\documentclass[aps,twocolumn,amsmath,amssymb,floatfix,showpacs,
superscriptaddress]{revtex4}

\usepackage[dvips]{graphics}
\usepackage{color}
\definecolor{gold}{rgb}{0.85,0.66,0}
\definecolor{dblue}{rgb}{0.3,0.7,0}

\begin{document}

\title{\textcolor{dblue}{Magneto-transport in a mesoscopic ring
with Rashba and Dresselhaus spin-orbit interactions}}

\author{Santanu K. Maiti}

\email{santanu.maiti@saha.ac.in}

\affiliation{Theoretical Condensed Matter Physics Division, Saha
Institute of Nuclear Physics, Sector-I, Block-AF, Bidhannagar,
Kolkata-700 064, India}

\affiliation{Department of Physics, Narasinha Dutt College, 129
Belilious Road, Howrah-711 101, India}

\author{Moumita Dey}

\affiliation{Theoretical Condensed Matter Physics Division, Saha
Institute of Nuclear Physics, Sector-I, Block-AF, Bidhannagar,
Kolkata-700 064, India}

\author{Shreekantha Sil}

\affiliation{Department of Physics, Visva-Bharati, Santiniketan, West
Bengal-731 235, India}

\author{Arunava Chakrabarti}

\affiliation{Department of Physics, University of Kalyani, Kalyani,
West Bengal-741 235, India}

\author{S. N. Karmakar}

\affiliation{Theoretical Condensed Matter Physics Division, Saha
Institute of Nuclear Physics, Sector-I, Block-AF, Bidhannagar,
Kolkata-700 064, India}

\begin{abstract}
Electronic transport in a one-dimensional mesoscopic ring 
threaded by a magnetic flux is studied in presence of Rashba and 
Dresselhaus spin-orbit interactions. A completely analytical technique 
within a tight-binding formalism unveils the spin-split bands in 
presence of the spin-orbit interactions and leads to a method of 
determining the strength of the Dresselhaus interaction. In addition 
to this, the persistent currents for ordered and disordered rings 
have been investigated numerically. It is observed that, the presence 
of the spin-orbit interaction, in general, leads to an enhanced 
amplitude of the persistent current. Numerical results corroborate 
the respective analytical findings.
\end{abstract}


\maketitle

\section{Introduction}

The spin-orbit (SO) interaction in mesoscopic and nano-scale semiconductor 
structures has been at the center stage of research in condensed matter 
theory and device engineering in recent times. The principal reason is 
its potential application in spintronics, where the possibility of 
manipulating and controlling the spin of the electron rather than its 
charge, plays the all important role~\cite{zutic,datta,ando,ding,bellucci,
aharony1,aharony2,citro,sheng,splett,meijer}. 

The spin-orbit fields in a solid are called the RSOI or the DSOI depending 
on whether the electric field originates from a structural inversion 
asymmetry or the bulk inversion asymmetry respectively~\cite{meier}.  
Quantum rings formed at the interface of two semiconducting materials 
are ideal candidates where the interplay of the two kinds of SOI might 
be observed. A quantum ring in a heterojunction is realized when a two 
dimensional gas of electrons is trapped in a quantum well due to the 
{\it band offset} at the interface of two different semiconducting 
materials. This {\it band offset} creates an electric field which may 
be described by a potential gradient normal to the interface~\cite{premper}. 
The potential at the interface is thus asymmetric, leading to the presence 
of a RSOI. On the other hand, at such interfaces, the bulk inversion 
symmetry is naturally broken.

In addition to this, the topology of such a ring gives rise to remarkable 
properties typical of low dimensional systems, for example, the persistent 
current. The phenomenon of persistent current in a conducting mesoscopic 
ring threaded by an Aharonov-Bohm (AB) flux $\phi$ has been established 
over many decades. The existence of discrete energy levels and large phase 
coherence length allow a non-decaying current upon the application of 
an external magnetic flux $\phi$. B\"{u}ttiker {\em et al.}~\cite{butt} 
first studied the behavior of persistent current in a metallic ring, 
and then many attempts have been made to explore the basic mechanisms 
of persistent current in mesoscopic ring and cylindrical 
systems~\cite{cheu1,alts,schm,ambe,ore,peeters,abra,bouz}. Later, several 
experiments~\cite{levy,chand,jari,deb,rabaud,mailly,blu} have also been 
performed to verify the existence of non-decaying current in these systems. 

Although the studies involving the mesoscopic rings have already generated 
a wealth of literature there is still need to look deeper into the problem 
\begin{figure}[ht]
{\centering \resizebox*{4.5cm}{2.5cm}{\includegraphics{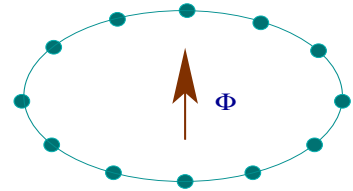}}\par}
\caption{(Color on-line) A mesoscopic ring threaded by an AB flux $\phi$.}
\label{ring}
\end{figure}
both from the point of view of fundamental physics and to resolve a few 
issues that have not yet been answered in an uncontroversial manner. In 
the present communication we undertake an in-depth analysis of the band 
structure of a mesoscopic ring in the presence of both the RSOI and the 
DSOI within a tight-binding (TB) formalism. We wish to look for a possible 
method of determining the strength of the DSOI in a 
simply connected mesoscopic ring. In recent past there have been a few 
experiments to measure the strength of the DSOI by optically monitoring 
the spin precession of the electrons~\cite{meier,studer}, through a 
measurement of the electrical conductance of narrow quantum wires defined 
in a 2DEG~\cite{scheid,jaya} or, by photogalvanic methods~\cite{yin}. 
Several studies already exist which deal with electron
spin states in mesoscopic rings in presence of both RSOI and DSOI and
the interplay between them~\cite{sheng}, the persistent current in
presence of SO interaction~\cite{splett} and the spin-Hall 
conductance~\cite{moca} using the continuum as well as the tight-binding
Hamiltonians. However, an analytical proposal to measure the strength 
of the DSOI, particularly within a tight-binding formalism, is still 
lacking, to the best of our knowledge.  

In the present work we consider a mesoscopic ring threaded by an AB flux, 
and work within a tight-binding approximation. 
Both the RSOI and the DSOI are included in the Hamiltonian. While looking 
for a way to estimate the strength of the DSOI, we work out a method to 
calculate the energy dispersion of the mesoscopic ring either with RSOI or 
with the DSOI, thus gaining a clean insight into the band structure. This 
method leads finally to the fact that by making the strengths of the RSOI 
and the DSOI equal, one comes across an absolute minimum in the conductance 
of the ring. As the RSOI is tuned, and hence estimated by controlling an 
external gate voltage, the estimation of the DSOI becomes obvious. The 
conductance is calculated via the Drude weight~\cite{kohn} numerically, 
and is found to support the analytical understanding.

In the second part, we show that the presence of the SO interaction leads 
to an enhancement of persistent current in one-dimensional rings. In 
addition to these, we present the detailed energy band structures and the 
oscillations of persistent current as the RSOI is varied to get a complete 
picture of the phenomena at the microscopic level. 

\section{The Model}

Let us start by referring to Fig.~\ref{ring}, where a mesoscopic ring is 
subject to an AB flux $\phi$ (measured in unit of the elementary flux 
quantum $\phi_0=ch/e$). Within a TB framework the Hamiltonian for such an 
$N$-site ring is~\cite{sheng,splett,moca} (and references therein), 
\begin{equation}
H = H_0 + H_{so}.
\label{equ1}
\end{equation}
Here,  
\begin{equation}
H_0 = \sum_n \mbox{\boldmath $c_n^{\dagger} \epsilon_0 c_n$} +
\sum_n \left(\mbox{\boldmath $c_n^{\dagger} t$} \,e^{i \theta} \mbox 
{\boldmath $c_{n+1}$} + h.c. \right)
\label{equ22} 
\end{equation}
and,
\begin{equation}
H_{so}= -\sum_n \left[\mbox{\boldmath $c_n^{\dag} t_{so}$} e^{i\theta} 
\mbox{\boldmath $c_{n+1}^{\dag}$} + h.c. \right]
\label{equ3}
\end{equation}
where,
\begin{eqnarray}
\mbox{\boldmath$t_{so}$} & = & i t_{Rso}\left(\mbox{\boldmath$\sigma_x$} 
\cos\varphi_{n,n+1} + \mbox{\boldmath$\sigma_y$} \sin\varphi_{n,n+1}
\right) \nonumber \\
& - & i t_{Dso} \left(\mbox{\boldmath$\sigma_y$} \cos\varphi_{n,n+1} 
+ \mbox{\boldmath$\sigma_x$} \sin\varphi_{n,n+1}\right).
\label{eq44}
\end{eqnarray}
$n=1$, $2$, $\dots$, $N$ is the site index along the azimuthal direction 
$\varphi$ of the ring. The other factors in Eqs.~\ref{equ22} and \ref{eq44}
are as follows.
\vskip 0.2cm
\noindent
\mbox{\boldmath $c_n$}=$\left(\begin{array}{c}
c_{n \uparrow} \\
c_{n \downarrow}\end{array}\right);$
\mbox{\boldmath $\epsilon_0$}=$\left(\begin{array}{cc}
\epsilon_0 & 0 \\
0 & \epsilon_0 \end{array}\right);$ 
\mbox{\boldmath $t$}=$t\left(\begin{array}{cc}
1 & 0 \\
0 & 1 \end{array}\right)$. \\
~\\
\noindent
Here $\epsilon_0$ is the site energy of each atomic site of the ring.
$t$ is the nearest-neighbor hopping integral and $\theta=2\pi \phi/N$
is the phase factor due to the AB flux $\phi$ threaded by the ring.
$t_{Rso}$ and $t_{Dso}$ are the isotropic nearest-neighbor transfer 
integrals which measure the strengths of Rashba and Dresselhaus SO 
couplings, respectively, and $\varphi_{n,n+1}=\left(\varphi_n+
\varphi_{n+1}\right)/2$, where $\varphi_n=2\pi(n-1)/N$. 
\mbox{\boldmath $\sigma_x$} and \mbox{\boldmath $\sigma_y$} are the 
Pauli spin matrices. $c_{n \sigma}^{\dagger}$ ($c_{n \sigma}$) is the 
creation (annihilation) operator of an electron at the site $n$ with 
spin $\sigma$ ($\uparrow,\downarrow$). Throughout the presentation we 
choose the units where $c=e=h=1$ and measure the SO coupling strength
in unit of $t$.

\section{Determination of the DSOI}

An elementary analysis of the effect of the SO interaction on the spectral 
and transport properties of the Hamiltonian may now be in order. To this 
end, we consider a mesoscopic ring with RSOI and DSOI, but without any AB 
flux threading the ring. The SO part of the Hamiltonian is then re-written 
as, 
\begin{equation}
H_{so}= -\sum_n \left[\mbox{\boldmath $c_n^{\dag} t_{so}$} 
\mbox{\boldmath $c_{n+1}^{\dag}$} + h.c. \right]
\label{equ99}
\end{equation}
which, on expansion becomes, 
\begin{equation}
H_{so}  =  -\mbox{\boldmath $t$} \sum_{n} 
\mbox{\boldmath $c_n^{\dag}c_{n+1}$} - i \sum_{n} 
\mbox{\boldmath $c_n^\dag$} \left(\begin{array}{cc}
0 & \mathcal P \nonumber \\
\mathcal Q & 0 \end{array}\right)
\mbox{\boldmath $c_{n+1}$} + h.c.
\end{equation}
where, 
\begin{eqnarray}
\mathcal P & = & \left(t_{Rso} + i t_{Dso}\right) \cos\varphi_{n,n+1} 
\nonumber \\
& - & i \left(t_{Rso} - i t_{Dso}\right) \sin\varphi_{n,n+1} \nonumber \\
\mathcal Q & = & \left(t_{Rso} - i t_{Dso}\right) \cos\varphi_{n,n+1}
\nonumber \\ 
 & + & i \left(t_{Rso} + i t_{Dso}\right) \sin\varphi_{n,n+1} 
\end{eqnarray}
A straightforward algebra helps us to recast the above Hamiltonian in 
Eq.~\ref{equ99} in the form,
\begin{equation}
H_{so} = -i \sqrt{t_{Rso}^2 + t_{Dso}^2} \sum_{n} \mathcal F_{n,n+1} \,
\mbox{\boldmath $c_n^{\dagger} {M} c_{n+1}$} + h.c.
\label{hso}
\end{equation} 
where, 
\begin{equation}
\mathcal F_{n,n+1} = \sqrt{1 - \sin(2 \xi) \sin(2 \varphi_{n,n+1})}
\label{factor}
\end{equation}
and, 
\begin{eqnarray}
\mbox {\boldmath $M$} & = & \left(\begin{array}{cc}
0 & e^{-i \chi_{n,n+1}} \nonumber \\
e^{i\chi_{n,n+1}} & 0 \end{array}\right)
\end{eqnarray}
with $\xi = \tan^{-1}\left(\frac{\displaystyle t_{Dso}}
{\displaystyle t_{Rso}}\right)$ and, \\
$\tan \chi_{n,n+1} = \frac{\displaystyle \sin(\varphi_{n,n+1}-\xi)}
{\displaystyle \cos(\varphi_{n,n+1}+\xi)}$.
\vskip 0.1cm
\noindent
We now analyze Eq.~\ref{hso} for two different cases. 

\vskip 0.1cm
\noindent
$\bullet$ {\bf\underline {Ring with RSOI only:}} In this 
limit Eq.~\ref{hso} leads to the full Hamiltonian of a ring with $N$ sites. 
With $\epsilon_0 = 0$, and with RSOI alone, this reads, 
\begin{eqnarray}
H_{R} &=& -t \sum_{n} \left(\mbox{\boldmath $c_n^{\dag}c_{n+1}$} 
+ \mbox{\boldmath $c_{n+1}^{\dag}c_{n}$}\right) \nonumber \\
& - & i \sum_{n} \mbox{\boldmath $c_n^\dag$} \left(\begin{array}{cc}
0 & e^{-i\varphi_{n,n+1}} \nonumber \\
e^{i\varphi_{n,n+1}} & 0 \end{array}\right)
\mbox{\boldmath $c_{n+1}$} + h.c.
\label{rashba}
\end{eqnarray}
To get the energy dispersion relations analytically, we define a unitary 
operator 
\begin{eqnarray}
\mbox{\boldmath $\mathcal U_n$} = \frac{1}{\sqrt{2}} \left(\begin{array}{cc}
1 & -1 \nonumber \\
e^{2\pi (n-1/2)i/N} & e^{2\pi (n-1/2)i/N} \end{array} \right)
\end{eqnarray}
which transforms the old operators \mbox{\boldmath $c_n$} to a set of 
new operators 
\mbox{\boldmath $\tilde {c_n} = \mathcal U_n^\dag c_n$}.
With the operators defined in the new basis, and using the discrete 
Fourier transform,
\begin{equation}
\mbox{\boldmath $\tilde c_n$} = \frac{1}{\sqrt{N}} \sum_{k_m} e^{ik_ma} 
\mbox{\boldmath $c_k$}
\end{equation}
where, $a$ is the lattice spacing (and set equal to unity in the 
subsequent discussion), one obtains explicit expressions of eigenvalues 
as, 
\begin{eqnarray}
E_{k_m,+} & = & -2t \cos(\pi/N) \cos(k_m+\frac{\pi}{N}) \nonumber \\
& + & 2 \sin(k_m+\frac{\pi}{N}) \sqrt{t^2 \sin^2 (\frac{\pi}{N}) + 
t_{Rso}^2} 
\end{eqnarray}
and,
\begin{eqnarray}
E_{k_m,-} & = & -2t \cos(\pi/N) \cos(k_m+\frac{\pi}{N}) \nonumber \\
& - & 2 \sin(k_m+\frac{\pi}{N})\sqrt{t^2 \sin^2(\frac{\pi}{N}) + 
t_{Rso}^2} 
\end{eqnarray}
where, $k_m=2m\pi/N$. In presence of a magnetic flux $k_m$ is replaced 
by $(k_m+\theta)$.
 
A similar analysis can be done to extract the eigenvalues for the 
dispersion relations of a ring with DSOI only, that is, with RSOI set equal 
to zero. However, in presence of both the RSOI and DSOI one has to resort 
to numerical methods for calculating the eigenvalues and eigenstates.

\vskip 0.1cm
\noindent
$\bullet$ {\bf\underline {Analytical argument for an estimation of DSOI:}} 
The expression of $\mathcal F$ (Eq.~\ref{factor}) plays a key-role in 
providing a method of estimation of the strength of the DSOI.
In the expression of $\mathcal F$, the presence of the term 
$\sin(2 \xi) \sin(2 \varphi_{n,n+1})$ under the square root `generates' 
an effective site-dependent hopping integral in the Hamiltonian. This 
site dependence of the hopping integral enters through the 
$\sin(2 \varphi_{n,n+1})$ term, and is responsible for scattering as the 
electron circulates along the arm of the ring. The scattering reduces 
the conductivity (the Drude weight). From the expression of the factor 
$\mathcal F_{n,n+1}$, we see that, in the absence of $t_{Dso}$, $\xi = 0$, 
and the the effective hopping integral becomes site-independent. The 
transport is ballistic in this case. On the other hand, when 
$t_{Dso}=t_{Rso}$ the {\it fluctuation} part under the square-root is 
maximum. In this case $\sin{2 \xi} = 1$. The site dependent scattering 
is strongest in this case, and should produce a minimum in the conductivity. 
This is precisely what we observe in the numerical calculations.

\vskip 0.1cm
\noindent
$\bullet$ {\bf\underline {Numerical argument for an estimation of DSOI:}} 
We now present numerical results in support of this argument, and in 
addition to this, on the effect of SO interaction on the persistent 
\begin{figure}[ht]
{\centering \resizebox*{7.5cm}{5cm}{\includegraphics{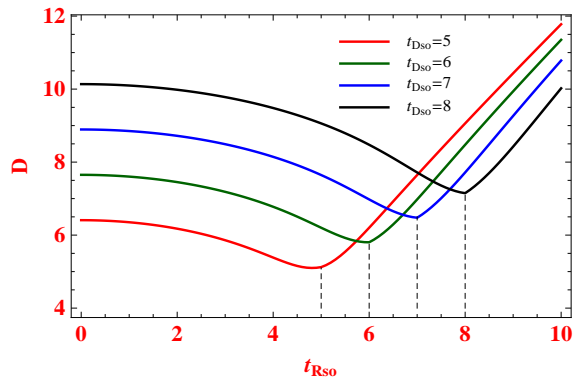}}\par}
\caption{(Color on-line) Drude weight $D$ (in unit of $N/4\pi^2$) as 
a function of Rashba SO coupling strength $t_{Rso}$ for a $40$-site 
ordered ring in the half-filled case considering different values of
$t_{Dso}$. $D$ reaches a minimum when $t_{Dso}=t_{Rso}$.}
\label{drude}
\end{figure}
current in such rings. While the results in the former case completely 
agrees with the predictions from the analytical calculations, the latter 
results exhibit remarkable variations including a marked enhancement in 
the amplitude of the persistent current in the presence of RSOI alone.
\begin{figure*}[ht]
{\centering \resizebox*{15cm}{9cm}{\includegraphics{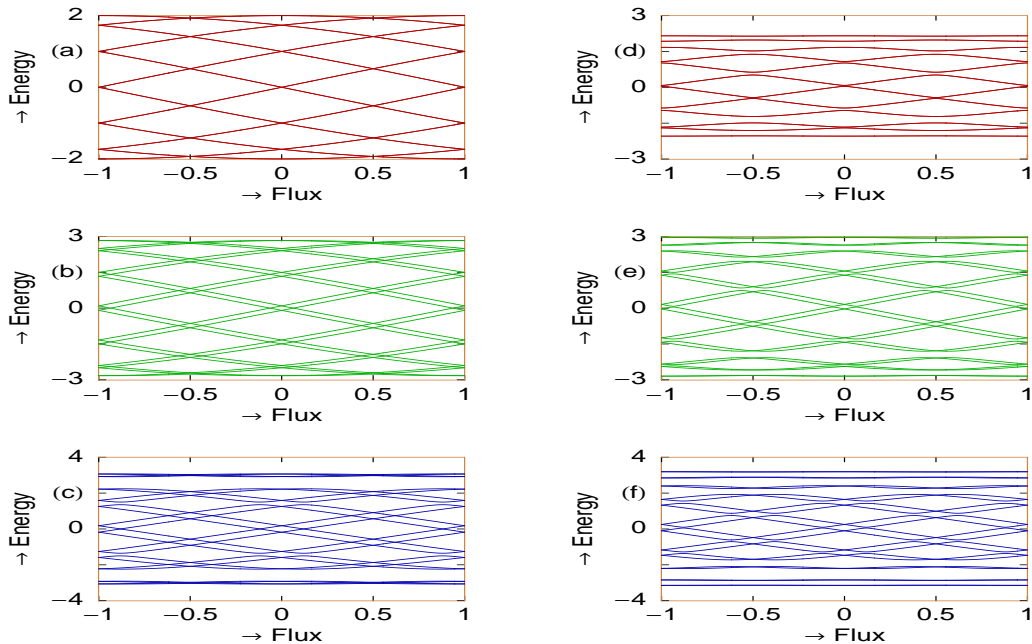}}\par}
\caption{(Color on-line) $E$-$\phi$ curves of a $12$-site ring, where the 
$1$st and $2$nd columns correspond to the results for the ordered ($W=0$) 
and disordered ($W=1$) cases, respectively. The red, green and blue lines 
correspond to $t_{Rso}=t_{Dso}=0$; $t_{Rso}=1$, $t_{Dso}=0$ and $t_{Rso}=1$, 
$t_{Dso}=0.5$, respectively.}
\label{energy}
\end{figure*}

We obtain numerical results for the conductance of a ring with 
different values of $t_{Rso}$ and $t_{Dso}$ by calculating the Drude 
weight $D$ in accordance with the idea originally put forward by 
Kohn~\cite{kohn}. The Drude weight for the ring is given by the relation,
\begin{equation}
D=\left . \frac{N}{4\pi^2} \left(\frac{\partial{^2E_0(\phi)}}
{\partial{\phi}^{2}}\right) \right|_{\phi \rightarrow 0}
\label{equ33}
\end{equation}
where, $N$ gives total number of atomic sites in the ring. Kohn
has shown that for an insulating system $D$ decays exponentially
to zero, while it becomes finite for a conducting system.

We have investigated the variation of the Drude weight $D$ as a function of 
the strength of the RSOI for an ordered ring at half-filling. The results are 
depicted in Fig.~\ref{drude}. A remarkable feature of the results presented 
is that, the conductivity (Drude weight) exhibits an absolute minimum 
whenever the strength of the DSOI becomes equal to the strength of the RSOI. 
This has been verified by choosing different values of $t_{Dso}$ for a 
$40$-site ordered ring and with electron number $N_e=40$. In each case, the 
conductance minimum appears exactly at $t_{Rso} = t_{Dso}$ confirming, as 
argued earlier, that a clear estimate of the strength of the DSOI may be 
obtained by noting the conductance minimum of a mesoscopic ring formed at 
the heterojuncion of two semiconduncting materials. As the strength of the 
RSOI is controlled by applying suitable gate voltage, the measurement of 
the DSOI becomes obvious. 

Before we end this section we would like to point our that the conclusion 
regarding the minimum in the Drude weight remains independent of the 
numerical values
chosen for $t_{Rso}$, and will occur whenever $t_{Dso}=t_{Rso}$.  However,
the question is about the detectability of this ``minimum". In this regard,
we would like to emphasize that observing the minimum numerically depends
strongly on the size of the ring (that is, the number of atoms taken in the
ring). One can bring down the minimum in the Drude weight to any desirable
value of $t_{Dso}$ by appropriately increasing the size of the ring. For
example, though in the paper we have presented results for a $40$-site ring
with the minima beginning at $t_{Rso}=5$, we have checked that the minimum
can be brought down to say, $t_{Rso}=0.4$ by choosing a ring with $160$ 
sites.

\section{Enhancement of persistent current}

At absolute zero temperature ($T=0$K), the persistent current in the
ring described with fixed number of electrons $N_e$ is determined by,
\begin{equation}
I\left(\phi\right)=-\frac{\partial E_0(\phi)}{\partial \phi}
\label{equ2}
\end{equation}
where, $E_0(\phi)$ is the ground state energy. We compute this quantity
to understand unambiguously the role of the RSOI interaction alone 
on persistent current.

Before presenting the results for $I(\phi)$, to make the present 
communication a self contained study, we first take a look at the energy 
spectrum of both an ordered and a disordered ring with and without the 
SO interactions, as the flux through the ring is varied. 
In Fig.~\ref{energy} the flux dependent spectra are shown for a $12$-site 
ordered ring and a randomly disordered one (with diagonal disorder) in the 
left and the right columns respectively. Clearly, disorder destroys the 
band crossings observed in the ordered case. The presence of the RSOI and 
the DSOI also lifts the degeneracy and opens up gaps towards the edges 
of the spectrum.

\begin{figure}[ht]
{\centering \resizebox*{7.5cm}{4.5cm}
{\includegraphics{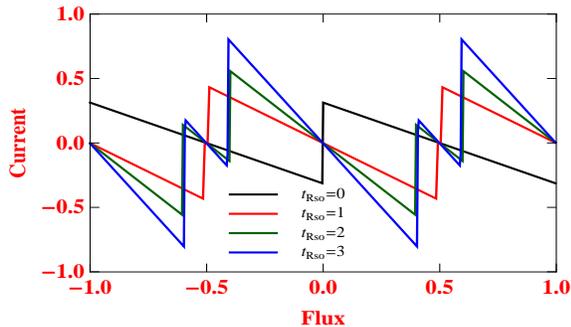}}\par}
\caption{(Color on-line) Current-flux characteristics of a $80$-site 
ordered ($W=0$) half-filled ring for different values of $t_{Rso}$ when 
$t_{Dso}$ is set at $0$.}
\label{current}
\end{figure}

\vskip 0.1cm
\noindent
$\bullet$ {\bf\underline{An ordered ring:}} In Fig.~\ref{current} we 
examine the effect of the RSOI on the persistent current of an ordered 
ring with $80$ sites. The DSOI is set equal to zero. We have examined 
both the non-half-filled and half-filled band cases, but present 
results for the latter only to save space. With increasing strength 
of the RSOI the persistent current exhibits a trend of an increase in 
its amplitude. Local phase reversals take place together with the 
appearance of kinks in the current-flux diagrams which are however, 
not unexpected even without the RSOI, and are results of the band 
crossings observed in the spectra of such rings. The amplitude of 
the persistent current at a specific value of the magnetic flux is of 
course not predictable in any simple manner, and is found to be highly 
sensitive to the number of electrons $N_e$ (i.e., the filling factor). 
Issues related to the dependence of the persistent current on the 
filling factor have been elaborately discussed by Splettstoesser 
{\em et al.}~\cite{splett}.

The persistent current in an ordered ring also exhibits interesting 
oscillations in its amplitude as the RSOI is varied keeping the magnetic 
flux fixed at a particular value. The oscillations persist irrespective 
\begin{figure}[ht]
{\centering \resizebox*{7.5cm}{4.5cm}
{\includegraphics{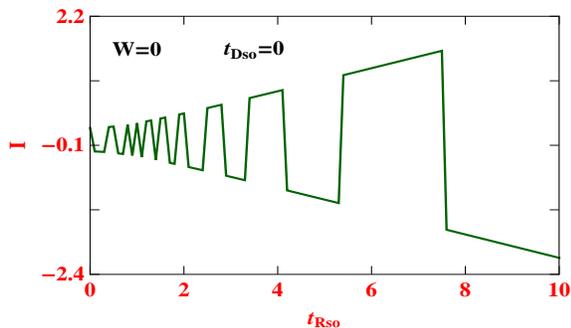}}\par}
\caption{(Color on-line) Persistent current at a particular AB flux 
($\phi=0.25$) as a function Rashba SO interaction strength for an ordered 
($W=0$) half-filled ring with $N=60$ when $t_{Dso}$ is set to zero.}
\label{typicurr3}
\end{figure}
of the band-filling factor $N_e$, with or without the presence of the 
DSOI. {\textcolor{red}{In Fig.~\ref{typicurr3} the oscillating nature 
of persistent current is presented for a $60$-site ordered ring in the 
half-filled band case when $\phi$ is set at $\phi_0/4$.}} The current 
exhibits oscillations with growing amplitude as the strength of the 
RSOI is increased.  

\begin{figure}[ht]
{\centering \resizebox*{7.5cm}{4.5cm}
{\includegraphics{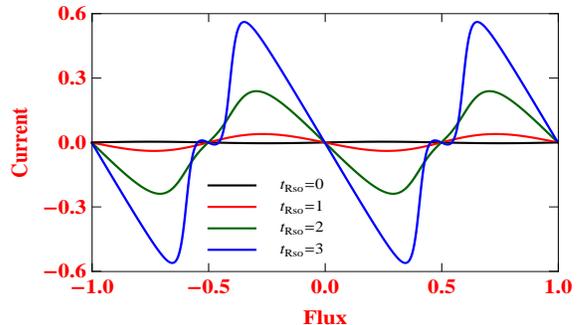}}\par}
\caption{(Color on-line) Current-flux characteristics of a $80$-site 
disordered ($W=2$) half-filled ring for different values of $t_{Rso}$ 
when $t_{Dso}$ is fixed at $0$.}
\label{currentdisorder1}
\end{figure}

\vskip 0.1cm
\noindent
$\bullet$ {\bf\underline {A disordered ring:}} We now present the results 
for a disordered ring of $80$ sites in Fig.~\ref{currentdisorder1}. 
Disorder is introduced via a random distribution (width $W =2$) of the 
values of the on-site potentials (diagonal disorder), and results averaged 
over sixty disorder configurations have been presented. The DSOI remains 
zero. Without any spin-orbit interaction, disorder completely suppresses 
the persistent current (an effect of the localization of the electronic 
states in the ring), as it is observed in Fig.~\ref{currentdisorder1} 
(black curve). With the introduction of the RSOI, the current starts 
increasing, and for $t_{Rso}=3$ (blue curve), increases significantly, 
attaining a magnitude comparable to that in a perfectly ordered ring. 
It is to be noted that the strength of the RSOI is strongly dependent 
on gate voltage. An enhancement 
of the persistent current in the presence of disorder can be achieved
even with much lower values of the RSOI parameter compared to what have
\begin{figure}[ht]
{\centering \resizebox*{7.5cm}{4.5cm}
{\includegraphics{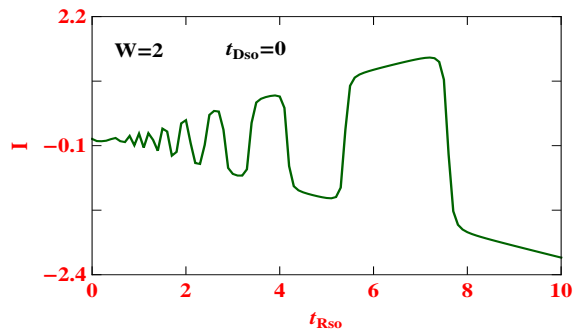}}\par}
\caption{(Color on-line) Persistent current at a particular AB flux 
($\phi=0.25$) as a function Rashba SO interaction strength for a 
disordered ($W=2$) half-filled ring with $N=60$ when $t_{Dso}$ is set
to zero.} 
\label{typicalcurrentdisorder}
\end{figure}
already been presented in the figures. To achieve this one needs 
to increase the size of the mesoscopic ring. We have checked this with
a $100$-site ring where even with $t_{Rso}=0.5$ the current increases 
by an order of magnitude compared to the case when $t_{Rso}=0$. However,
we present the results using a somewhat larger values of $t_{Rso}$ for a
better viewing of the results. Similar observations are made by setting 
$t_{Rso}=0$ and varying $t_{Dso}$.  

Disorder introduces quantum interference which leads to localization of 
the electronic states. RSOI, on the other hand, introduces spin flip 
scattering in the system, 
which can destroy quantum interference effect, leading to a possible 
delocalization of the electronic states. This leads to an enhancement 
of the persistent current in the presence of disorder. The competition 
between the strength of disorder and the RSOI is also apparent in 
Fig.~\ref{typicalcurrentdisorder}. For small values of the RSOI, the 
disorder dominates. As the strength of the RSOI is increased, the spin 
flip scattering starts dominating over the quantum interference effect, 
and finally the oscillations become quite similar to that in a ballistic 
ring. As the SO interaction is a natural interaction for a quantum ring 
grafted at a heterojunction, we are thus tempted to propose that the 
spin-orbit interaction is responsible for an enhanced persistent current 
in such mesoscopic disordered rings. 

Before we end this section, we would like to mention that
the presence of DSOI alone leads to exactly similar results as expected, 
since the Rashba and the Dresselhaus Hamiltonians are related by a unitary
transformation. This does not change the physics. We have also computed the
persistent current in the presence of both the interactions. The amplitude
of the current does not increase significantly compared to the case where
only one interaction is present. However, the precise magnitude of the 
current is sensitive to the strength of the magnetic flux threading the 
ring. The observation remains valid even when the strengths of the Rashba
and Dresselhaus spin-orbit interactions are the same.

\section{Conclusion}

In conclusion, we have investigated the spectrum and the magnetic response 
of a tight-binding mesoscopic ring with Rashba and Dresselhaus spin-orbit 
interactions both analytically and numerically. Two principal results have 
been obtained and discussed. First, after an exact analysis of the spin 
dependent Hamiltonian we argue that a minimum in the conductance of the 
ring system when the DSOI equal the strength of the RSOI provides a method 
of estimating the strength of the former. Second, we present numerical 
results of the band structure of the ring system when it is threaded by 
an AB flux $\phi$. The effect of the spin-orbit terms on the 
energy bands are shown. This is followed by an exhaustive numerical 
calculation of the persistent current in ordered and disordered rings in 
the presence of the spin-orbit interactions. The result for the disordered 
ring exhibits a large persistent current, of the same order of magnitude 
as that in a perfectly ordered ring, in the presence of Rashba spin-orbit 
interaction.

\end{document}